\documentclass[pdflatex,sn-mathphys-num]{sn-jnl}

\usepackage{graphicx}
\usepackage{booktabs}
\usepackage{amsmath}
\providecommand{\allowdisplaybreaks}{}

\title{Detailed investigation on the Geant4 simulation of alpha particles in liquid scintillator detector}

\author[1,2]{Shiwen Xu}
\author*[1,2]{Shubing Liu}
\email{liusb@ihep.ac.cn}
\author*[1]{Zeyuan Yu}
\email{yuzy@ihep.ac.cn}
\affil[1]{Institute of High Energy Physics, Chinese Academy of Sciences, Beijing 100049, China}
\affil[2]{University of Chinese Academy of Sciences, Beijing 100049, China}

\date{}

\abstract{
\textbf{Purpose} Alpha particles from $^{214}$Po, $^{212}$Po, and $^{210}$Po are increasingly used for energy calibration in large liquid scintillator detectors. However, the Geant4 simulation of alpha particle energy deposition in liquid scintillators lacks detailed validation. This work presents a systematic investigation of the simulation framework and identifies the dominant sources of energy resolution fluctuations.

\textbf{Methods} Geant4 11.4.2 simulates alpha particles in a 4~m cube of LAB-based liquid scintillator, with Birks quenching applied per step. The nominal setup uses a modified Bragg-model minimum and a 0.001~mm production range cut whose energy conversion has a 0.5~keV lower bound. Threshold and electromagnetic-model dependence is studied, and Geant4-DNA examines keV-scale electron track structure.

\textbf{Results} Delta-electron-associated fluctuations dominate the simulated alpha response: across the tested explicit-delta settings, the variance of the quenched-response contribution from secondary tracks is 1.01--1.12 times the total response variance, with a negative primary--secondary covariance reducing the total. The simulation is compatible with the SNO+ alpha quenching curve shape within the current experimental uncertainties, giving $\chi^2$/ndf $\simeq 0.14$--0.17 for $k_B = 0.007$--0.010~g$\cdot$cm$^{-2}\cdot$MeV$^{-1}$. Geant4-DNA simulations reveal strong track curvature (tortuosity $\sim$6 at 200~eV, decreasing to $\sim$1.6 at 10~keV), motivating tests of the continuous column model for Birks quenching of keV-scale delta electrons.

\textbf{Conclusion} Delta electron fluctuations dominate the intrinsic energy resolution, with a smaller contribution from the alpha primary track. The delta electron threshold has a larger effect than the choice of electromagnetic model in this study; Penelope and Livermore agree to within $\sim$3\% in L/E, while Standard gives higher L/E, especially at low thresholds.
}


\begin{document}
\maketitle

\section{Introduction}

Alpha particles from the natural radioactive decay chains of $^{214}$Po (7.69~MeV), $^{212}$Po (8.78~MeV), and $^{210}$Po (5.30~MeV) are playing an increasingly important role in the energy calibration of large liquid scintillator (LS) detectors. Unlike artificial calibration sources (e.g., $^{68}$Ge, $^{241}$Am--Be), these alpha emitters are dissolved directly in the scintillator volume, providing a uniform and continuous calibration source. The alpha particles are mono-energetic and free from the energy absorption and light-blocking effects associated with capsule-based calibration sources. This makes them particularly attractive for monitoring the long-term stability of the energy scale and for studying the intrinsic energy resolution of the detector.

However, the accurate simulation of alpha particle energy deposition in LS is essential for interpreting these calibration signals. The energy loss of alpha particles in organic scintillators is dominated by two processes: (i) the continuous slowing down of the primary alpha track, and (ii) the production of energetic delta electrons that carry a fraction of the alpha's energy away from the primary track. The relative contribution of these two components, and their fluctuations, determine the intrinsic energy resolution of the alpha signal.

Several aspects of the Geant4 simulation framework for alpha particles require careful examination. First, in the tested no-delta configuration without a user-imposed step limiter, the alpha track is simulated in a single CSDA step, which yields a deterministic result with no event-by-event fluctuations. A finite step limit divides the alpha track into multiple steps and produces a nonzero response width. Second, delta-electron production is governed by the Bragg model's minimum energy and by the electron energy cut obtained from the production range cut; the specified lower bound of the range-to-energy conversion can clamp the latter. Third, the treatment of low-energy electrons affects the quenching calculation and is a major source of model dependence.

Recent desktop experiments~\cite{vonkrosigk, tretyak2009, vink2013} have measured the alpha quenching non-linearity in LAB-based scintillators and explored the role of delta electrons in alpha energy deposition. These measurements provide valuable constraints on the Birks quenching parameters. On the theoretical side, the ExcitonQuenching model~\cite{christensen} offers a physically motivated reaction--diffusion model based on the Blanc equation and track structure theory, separating the ion track into a high-density core and a delta-electron-dominated penumbra. However, this model was developed for plastic scintillators, and its direct application to liquid scintillators requires validation.

In this work, we present a systematic Geant4 simulation study of alpha particle energy deposition and light yield in LAB-based liquid scintillator. We investigate: (i) the role of the step limit and the origin of energy loss fluctuations in the absence of delta electrons, (ii) the energy partition between the alpha primary track and delta electrons and its dependence on the production threshold and Birks constant, (iii) the comparison with the SNO+ alpha quenching data as a shape cross-check, (iv) the energy resolution with delta electrons and its dependence on $k_B$ and the threshold, (v) the variance decomposition to identify the dominant fluctuation source, (vi) the dependence on the electromagnetic physics model (Standard, Penelope, Livermore), and (vii) the validity of the continuous column approximation for keV-scale delta electrons using Geant4-DNA track structure simulations. We also discuss the ExcitonQuenching model's prediction of the core vs.\ penumbra light partition and the challenges of simulating keV-scale delta electron quenching.

\section{Simulation Framework}

\subsection{Default Geant4 Behavior}

Geant4 11.4.2~\cite{geant4} is used. Unless otherwise stated, the simulation uses Geant4 standard electromagnetic processes (``Standard''); Penelope and Livermore are used as cross-checks in Sec.~\ref{sec:threshold}. The detector geometry is a 4~m cube of LAB-based liquid scintillator ($\rho = 0.859$~g/cm$^3$, C$_{10}$H$_{14}$). The Birks quenching formula is applied per step:

\begin{equation}
Q = \frac{\Delta E}{1 + k_B S_m}, \qquad S_m = \frac{1}{\rho}\frac{dE}{dx},
\label{eq:birks}
\end{equation}
where $dE/dx=\Delta E/\Delta x$ is the step-averaged linear stopping power and $S_m$ is the mass stopping power expressed in MeV$\cdot$cm$^2$/g. The Birks constant $k_B$ is quoted in g$\cdot$cm$^{-2}\cdot$MeV$^{-1}$. The event response $Q$ is summed over depositing tracks, and $\sigma_{\rm dep}=\operatorname{Std}(Q)/\langle Q\rangle$ excludes photon statistics and detector effects.

We first examine the no-delta configuration without explicit delta electron production. This is achieved by setting the production cut to 1~mm, which suppresses the production of secondary electrons. In this configuration, the alpha particle's energy is deposited entirely through the continuous energy loss of the primary track.

\begin{figure}[htbp]
\centering
\includegraphics[width=0.85\textwidth]{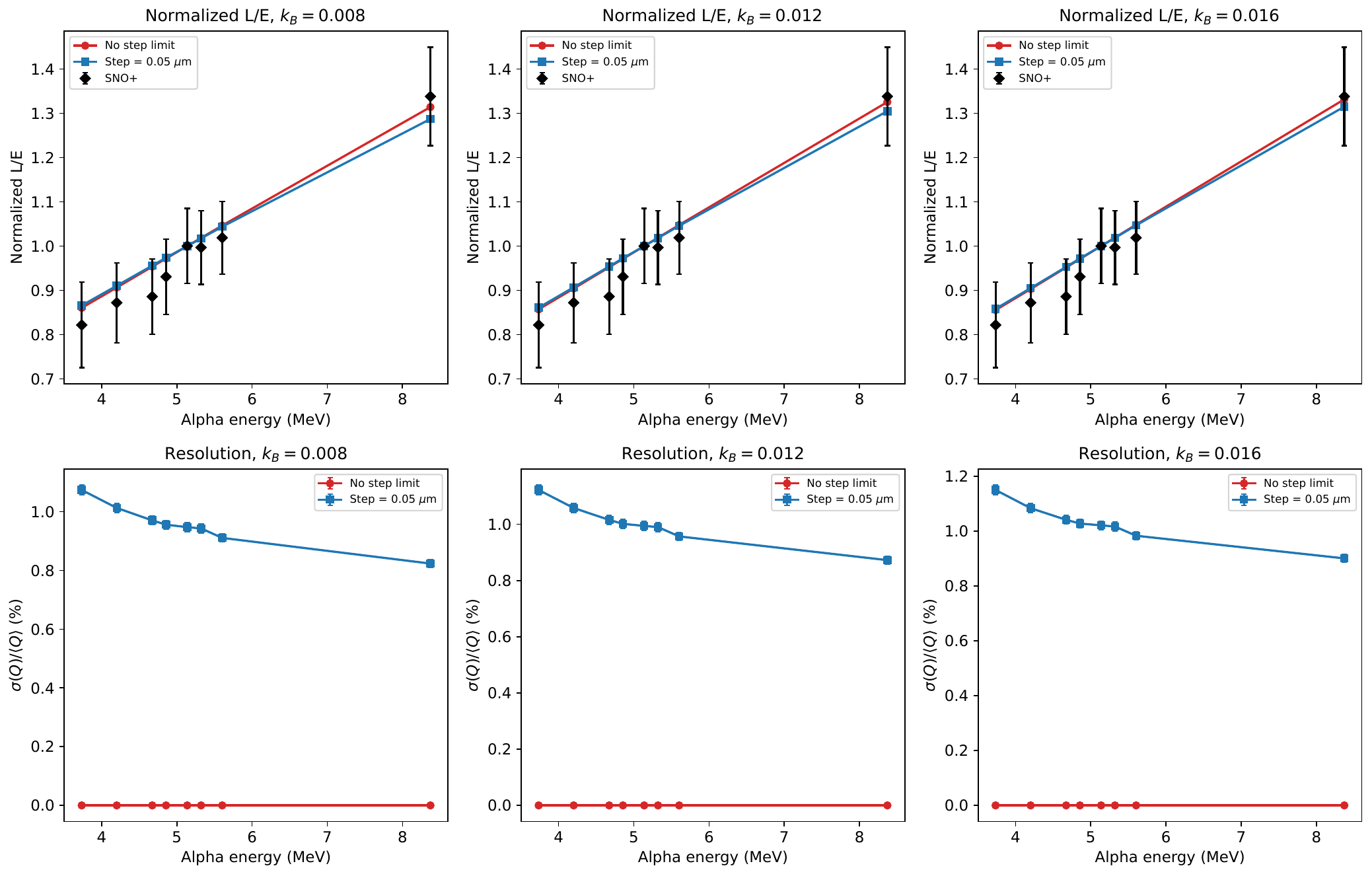}
\caption{Effect of step size on the no-delta alpha simulation at different Birks constants. \textbf{Top row}: normalized quenching curve (L/E normalized to 1 at 5.138~MeV), compared with SNO+ data (black diamonds). \textbf{Bottom row}: energy resolution. Without a step limit, the alpha is simulated in a single step, giving a deterministic result ($\sigma=0$). The finite-step curves use a max step of 0.05~$\mu$m; the result depends on the chosen finite limit. The normalized shape of the no-delta simulation deviates from the SNO+ data, especially at low energies.}
\label{fig:nodelta_stepsize}
\end{figure}

Without a user-imposed step limit, the tested no-delta configuration transports the entire alpha track in a single step. The Birks formula is applied once to the total energy deposition, yielding a completely deterministic result: the quenching curve L/E is a function of energy only, with zero event-by-event fluctuation ($\sigma = 0$). The L/E values at $k_B=0.008$ range from 0.069 at 3.735~MeV to 0.106 at 8.371~MeV. This is verified by 5000-event simulations where all events produce the identical quenched energy.

When a step limit is imposed, the alpha track is divided into multiple steps. Fluctuations in the energy deposition per step propagate through the nonlinear Birks formula, producing a non-zero energy resolution. At 5.138~MeV and $k_B=0.008$, step limits of 1~$\mu$m, 0.1~$\mu$m, and 0.05~$\mu$m give mean primary step counts of 42.2, 421.3, and 837.5, and resolutions of 1.478\%, 0.989\%, and 0.948\%, respectively. The L/E value at 0.05~$\mu$m is slightly lower than the single-step case (approximately 0.078 vs 0.080), because the nonlinear per-step Birks calculation gives a different result from the single-step calculation.

When delta electrons are suppressed (production cut = 1~mm), the energy deposition is dominated by the alpha primary track. Its fluctuations are strongly correlated with the track-averaged stopping power, as shown by an independent 9000-event sample for the 8.371~MeV case with a 0.05~$\mu$m step limit:

\begin{itemize}
\item Primary track length: $92.09 \pm 1.03~\mu$m
\item Average dE/dx: $E_\alpha/\ell_p=90.9\pm1.0$~MeV/mm, corresponding to $1058\pm12$~MeV$\cdot$cm$^2$/g
\item $\mathrm{Corr}(\text{dE/dx}, \text{QEDEP}) = -0.938$ (strong anti-correlation)
\end{itemize}

A single-effective-stopping-power approximation gives a relative sensitivity $f=(k_B S_m)/(1+k_B S_m)\approx0.89$, corresponding to an expected response width of approximately 0.95\%. The event-level result is $\sigma_{\rm dep}=0.849\pm0.012\%$; the difference reflects the full step history entering the Birks sum.

The no-delta resolution decreases with energy (from 1.07\% at 3.735~MeV to 0.82\% at 8.371~MeV). The no-delta quenching curve shape deviates more from the digitized SNO+ data, especially at low energies: the reduced $\chi^2$/ndf for the normalized shape is 0.24 without delta, compared to 0.16 with delta at $k_B=0.008$ (Sec.~\ref{sec:snocomparison}).

\begin{figure}[htbp]
\centering
\includegraphics[width=0.85\textwidth]{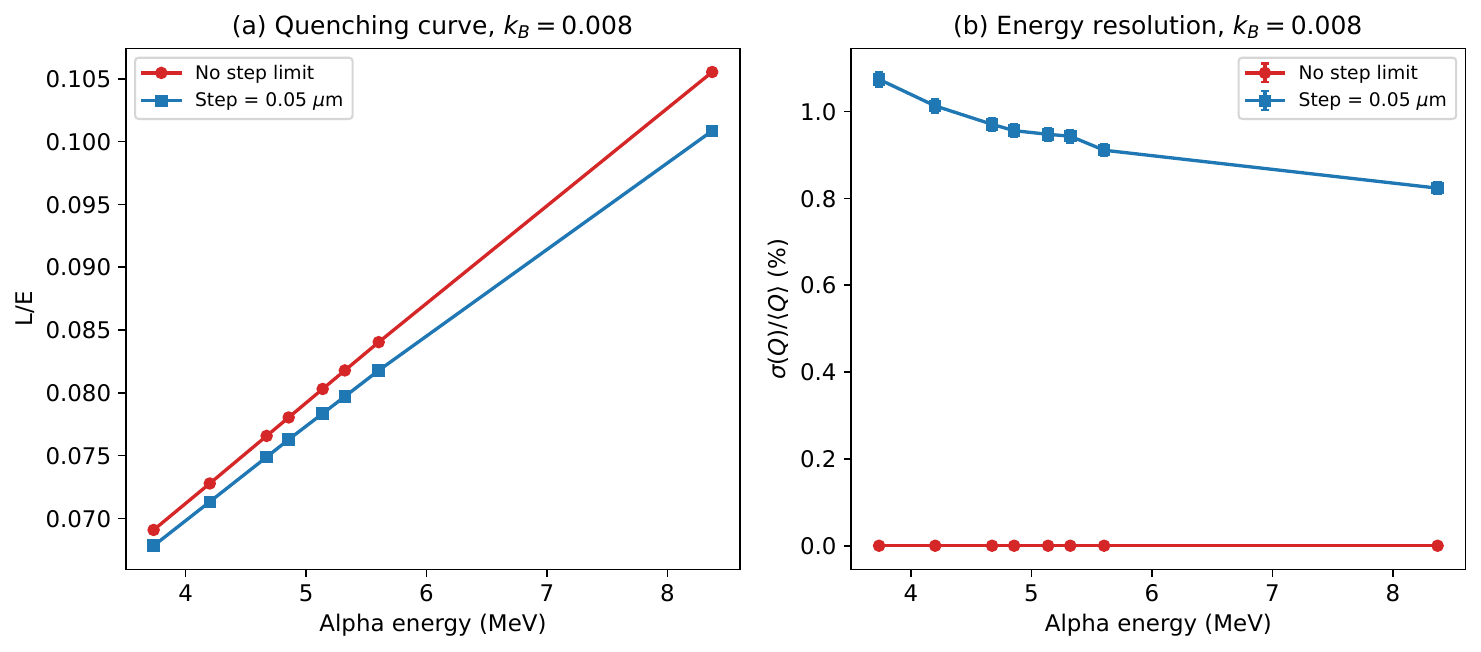}
\caption{Effect of the step limit on the no-delta alpha simulation. Without a step limit, the alpha is simulated in a single step, giving a deterministic result ($\sigma=0$). With a max step of 0.05~$\mu$m, the alpha is divided into multiple steps (518--1842 depending on energy), and the energy resolution appears. The L/E values are slightly lower with a step limit due to the nonlinear per-step Birks response.}
\label{fig:nodelta_stepsize_simple}
\end{figure}

\subsection{Delta Electron Production Threshold}

The production of delta electrons is controlled by the production cut (a range cut converted to energy) and the minimum kinetic energy parameter $E_{\text{min}}$ in the $G4BraggModel$ class. In the BraggIon regime, the effective threshold is the maximum of the converted cut, the material minimum cut, and the mass-scaled $E_{\text{min}}$.

\begin{figure}[htbp]
\centering
\includegraphics[width=0.65\textwidth]{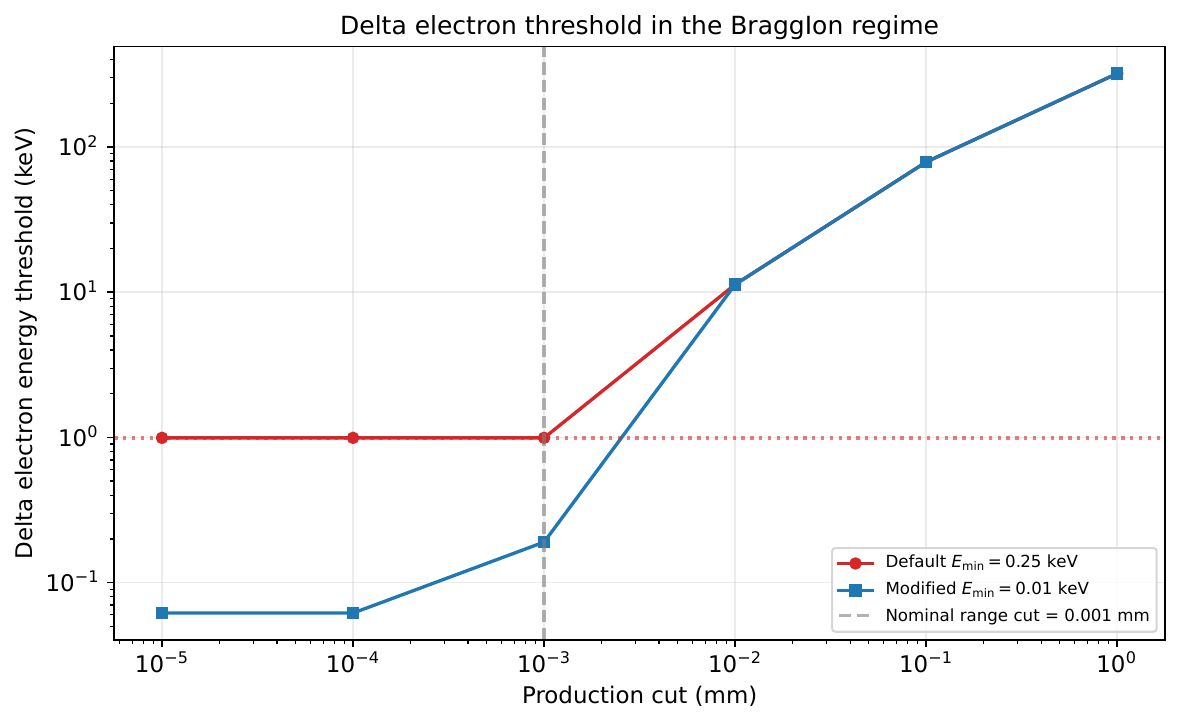}
\caption{Effective delta electron energy threshold in the BraggIon regime as a function of the production cut, using a conversion lower bound of 0.01~keV. With the default $E_{\rm min}=0.25$~keV, the threshold is bounded below by 0.99~keV. With the modified $E_{\rm min}=0.010$~keV, the threshold follows the converted cut down to the material minimum of 0.062~keV, reaching 0.19~keV at the nominal cut of 0.001~mm.}
\label{fig:threshold_control}
\end{figure}

The default Geant4 $G4BraggModel$ class has a hard-coded $E_{\text{min}} = 0.25$~keV, which for alpha particles ($m_\alpha/m_p = 3.97$) gives an effective threshold of $0.25 \times 3.97 = 0.99$~keV for delta electron production in the BraggIon regime. For production cuts smaller than $\sim$0.01~mm, the default $E_{\text{min}}$ dominates and the threshold is fixed at 0.99~keV, as shown in Fig.~\ref{fig:threshold_control}.

We modified this parameter to $E_{\text{min}} = 0.010$~keV, reducing the corresponding mass-scaled value to $\sim$0.04~keV. With this modification, the delta electron threshold in the BraggIon regime is determined by the production cut through the $G4VRangeToEnergyConverter$, subject to a material minimum of $\sim$0.062~keV. At the production cut of 0.001~mm, the threshold is $\sim$0.19~keV for sE$_{\rm min}=0.01$~keV and 0.5~keV for the nominal response setting sE$_{\rm min}=0.5$~keV.

\section{Results with Delta Electrons}
\label{sec:results}

\subsection{Delta Electron Energy and Multiplicity Distributions}

The production of delta electrons is controlled in part by sE$_{\rm min}$, which sets the lower bound for the range-to-energy conversion. We study the delta electron properties at four settings (0.1, 0.25, 0.5, and 1.0~keV) for 8.371~MeV alpha particles, counting each direct electron once at creation.

\begin{figure}[htbp]
\centering
\includegraphics[width=0.85\textwidth]{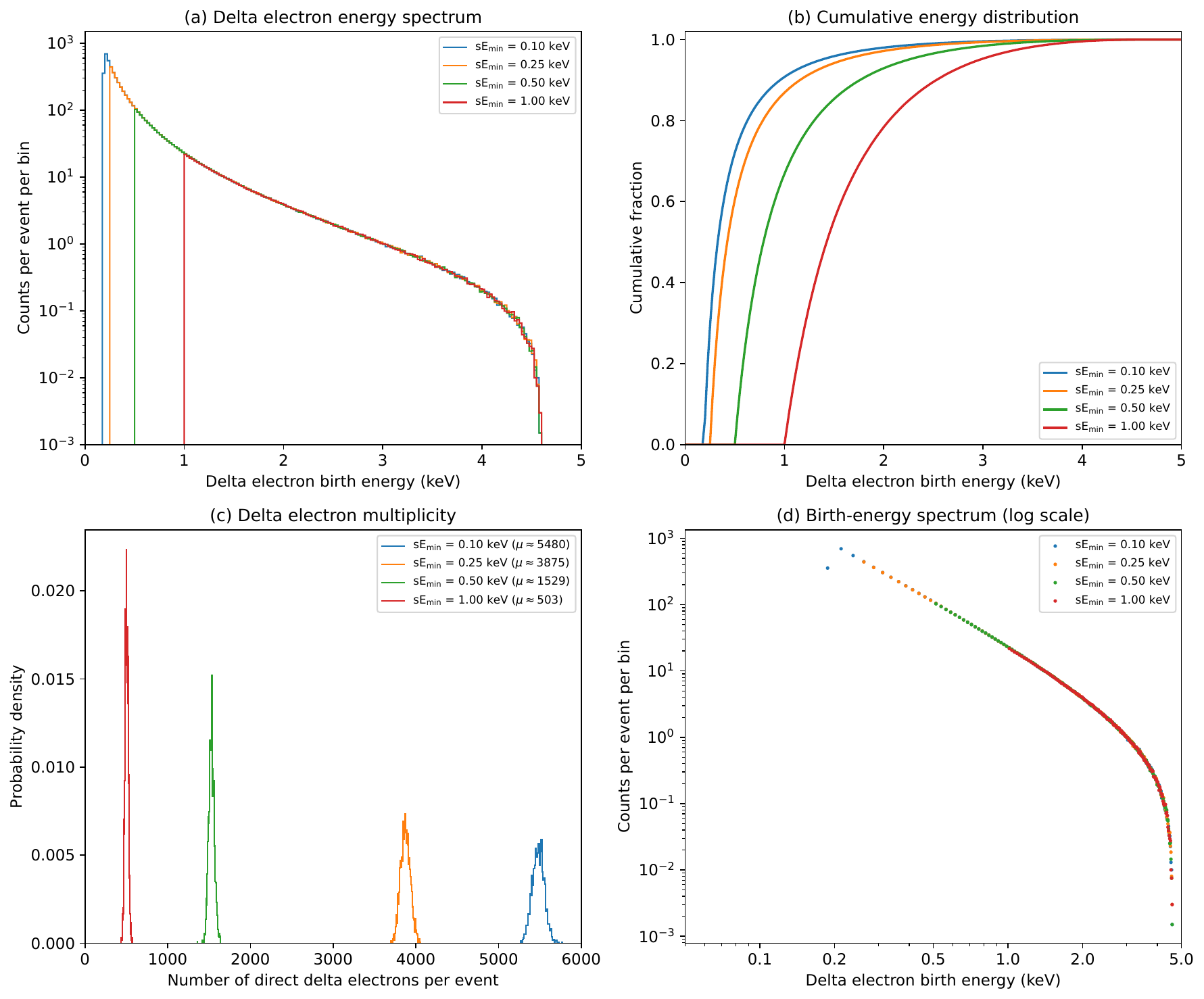}
\caption{Delta electron properties at different range-to-energy conversion lower bounds for 8.371~MeV alpha particles. (a) Birth-energy spectrum per event. (b) Cumulative distribution. (c) Unique direct-electron multiplicity. (d) Log--log view of the birth-energy spectrum.}
\label{fig:deltathresholds}
\end{figure}

The energy spectrum (Fig.~\ref{fig:deltathresholds}a,d) falls steeply with electron energy up to the kinematic limit $T_{\max} = 4 m_e E_\alpha / m_\alpha \approx 4.6$~keV for 8.371~MeV alpha particles. The cumulative distribution (Fig.~\ref{fig:deltathresholds}b) depends on the production setting, with a larger population of low-energy electrons at lower sE$_{\rm min}$. The multiplicity distribution (Fig.~\ref{fig:deltathresholds}c) is narrow compared with its mean, which increases as sE$_{\rm min}$ is lowered.

\subsection{Energy Partition at Different Thresholds and kB}

For an 8.371~MeV alpha particle, the energy is partitioned between the primary track and delta electrons. The delta electron contribution depends strongly on the production threshold sE$_{\rm min}$ and the Birks constant $k_B$. Table~\ref{tab:partition_thresholds} shows the energy partition and the quenched light yield fraction at three thresholds and three $k_B$ values.

\begin{table}[htbp]
\centering
\caption{Energy partition for 8.371~MeV alpha particles at different thresholds and Birks constants. $N_\delta$ is the mean unique direct-electron count. $p_{\rm edep}$ and $s_{\rm edep}$ are the energy deposited in the primary and all secondary tracks; $p_{\rm frac}$ and $s_{\rm frac}$ are their fractions of the total quenched light yield.}
\label{tab:partition_thresholds}
\begin{tabular}{c|c|c|c|c|c|c|c}
\hline
sE$_{\rm min}$ & $k_B$ & $N_\delta$ & $p_{\rm edep}$ & $s_{\rm edep}$ & $L/E$ & $p_{\rm frac}$ & $s_{\rm frac}$ \\
(keV) & & & (MeV) & (MeV) & & & \\
\hline
0.25 & 0.008 & 3875 & 6.033 & 2.338 & 0.2096 & 0.402 & 0.598 \\
0.25 & 0.012 & 3875 & 6.033 & 2.338 & 0.1591 & 0.380 & 0.620 \\
0.25 & 0.016 & 3875 & 6.033 & 2.338 & 0.1287 & 0.367 & 0.633 \\
\hline
0.5 & 0.008 & 1529 & 6.844 & 1.527 & 0.1766 & 0.524 & 0.476 \\
0.5 & 0.012 & 1529 & 6.844 & 1.527 & 0.1320 & 0.497 & 0.503 \\
0.5 & 0.016 & 1529 & 6.844 & 1.527 & 0.1059 & 0.481 & 0.519 \\
\hline
1.0 & 0.008 & 503 & 7.550 & 0.821 & 0.1454 & 0.666 & 0.334 \\
1.0 & 0.012 & 503 & 7.550 & 0.821 & 0.1070 & 0.637 & 0.363 \\
1.0 & 0.016 & 503 & 7.550 & 0.821 & 0.0851 & 0.620 & 0.380 \\
\hline
\end{tabular}
\end{table}

At the nominal threshold sE$_{\rm min}=0.5$~keV, delta electrons carry 18\% of the deposited energy but contribute 48\% of the quenched light yield. This is because the alpha core's dE/dx ($\sim$2400~MeV$\cdot$cm$^2$/g) is much higher than the delta electrons' dE/dx ($\sim$50--150~MeV$\cdot$cm$^2$/g), leading to much stronger quenching of the primary track. As the threshold is lowered, more delta electrons are produced, and their contribution to the light yield increases: at sE$_{\rm min}=0.25$~keV, the delta fraction $s_{\rm frac}$ reaches 60\%. Conversely, at sE$_{\rm min}=1.0$~keV, the delta fraction drops to 33\%.

The $k_B$ dependence is also significant: increasing $k_B$ from 0.008 to 0.016 reduces the total L/E by $\sim$40\%, with the delta electron contribution becoming slightly more important (higher $s_{\rm frac}$) because the primary track is more strongly quenched.

\subsection{Comparison with SNO+ Data}
\label{sec:snocomparison}

We compare the simulated alpha quenching curve with the LAB1 sample of the small-scale measurements by von Krosigk et al.\ 2016~\cite{vonkrosigk}, which measured alpha particle light output in LAB-based scintillator (LAB + 2~g/L PPO + 15~mg/L bis-MSB). The points were digitized from Fig.~6.30(a) of von Krosigk's thesis~\cite{vonkrosigkthesis}. The shape of the quenching curve, rather than its absolute normalization, is used for comparison to avoid the uncertainty in the electron light yield.

\begin{figure}[htbp]
\centering
\includegraphics[width=0.85\textwidth]{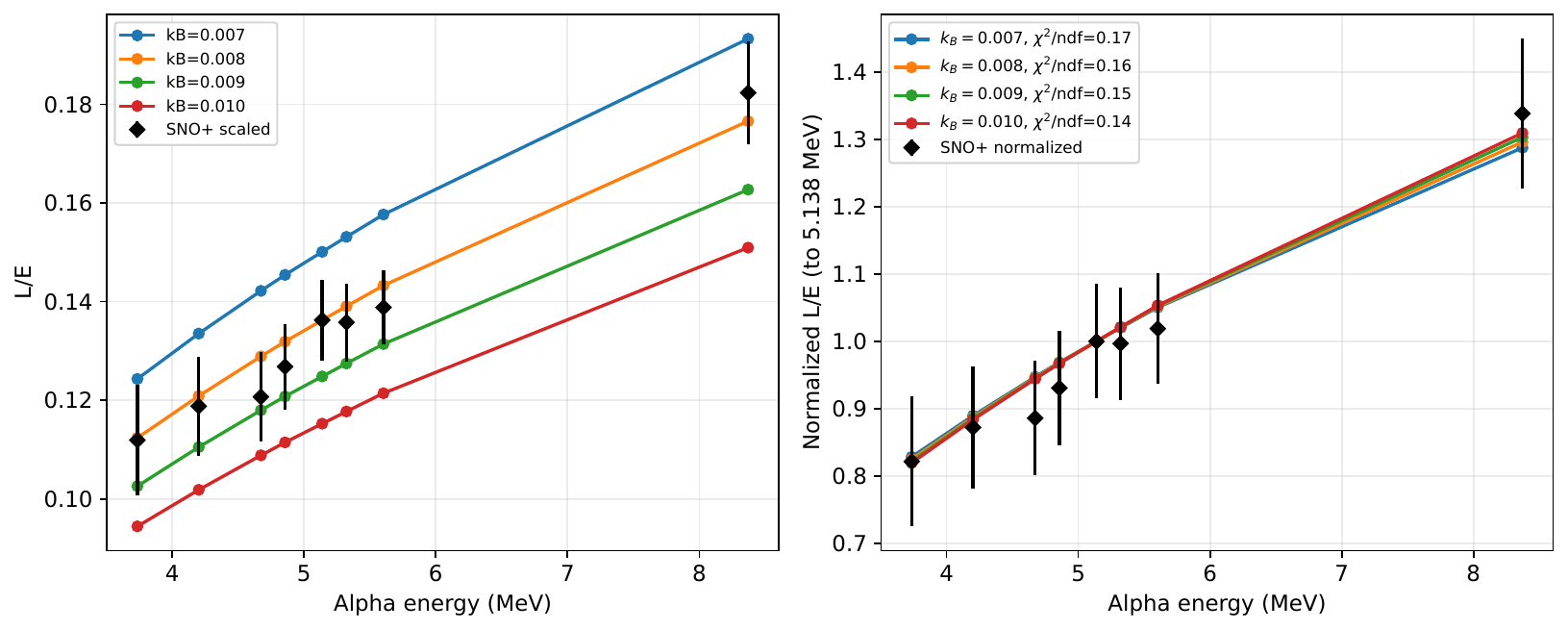}
\caption{Comparison of simulated and experimental alpha quenching curves at the nominal threshold sE$_{\rm min}=0.5$~keV. Left: absolute L/E, with SNO+ data scaled to match the simulation at 5.138~MeV. Right: normalized shape, with $\chi^2$/ndf values for each $k_B$ shown in the legend.}
\label{fig:sno}
\end{figure}

We perform a systematic scan of the Birks constant $k_B$ from 0.007 to 0.010~g$\cdot$cm$^{-2}\cdot$MeV$^{-1}$ (Fig.~\ref{fig:sno}). The normalized shape is obtained by dividing all L/E values by the value at 5.138~MeV. The $\chi^2$ values below use the digitized SNO+ ratio uncertainties and include an approximate propagation of the common normalization uncertainty; correlations introduced by this normalization are otherwise neglected, so the values should be interpreted as a shape compatibility test rather than a precision fit.

\begin{table}[htbp]
\centering
\caption{$\chi^2$/ndf for different Birks constants at the nominal threshold sE$_{\rm min}=0.5$~keV.}
\label{tab:kbscan}
\begin{tabular}{c|c}
\hline
$k_B$ (g$\cdot$cm$^{-2}\cdot$MeV$^{-1}$) & $\chi^2$/ndf \\
\hline
0.007 & 0.171 \\
0.008 & 0.159 \\
0.009 & 0.150 \\
0.010 & 0.143 \\
\hline
\end{tabular}
\end{table}

The normalized SNO+ shape is compatible with all scanned values within the current experimental uncertainties, with a mild preference toward the upper end of the scan. The range $k_B = 0.007$--0.010~g$\cdot$cm$^{-2}\cdot$MeV$^{-1}$ is consistent with Birks constants reported for organic liquid scintillators (0.006--0.010~g$\cdot$cm$^{-2}\cdot$MeV$^{-1}$)~\cite{tretyak2009, vink2013}. We keep $k_B=0.008$ as the nominal value for the resolution study because it lies in the standard LS range and gives a representative shape-compatible result.

We also examine the effect of the threshold on the shape agreement. Lowering or raising sE$_{\rm min}$ changes the simulated shape and therefore the comparison with SNO+. Given the size of the SNO+ uncertainties and the neglected normalization correlations, we do not use this comparison alone to determine a unique threshold; instead, sE$_{\rm min}=0.5$~keV is treated as a nominal setting for studying the threshold dependence.

\subsection{Energy Resolution with Delta Electrons}

Using the nominal parameters (sE$_{\rm min}=0.5$~keV, $k_B=0.008$), the energy resolution with delta electrons is $\sigma_{\rm dep}=1.42\%$ at 5.138~MeV, worsening to 1.68\% at 3.735~MeV and improving to 1.22\% at 8.371~MeV. This is a factor of $\sim$1.5 worse than the no-delta case, consistent with the dominant delta electron contribution found in Sec.~\ref{sec:variance}.

The resolution depends on both $k_B$ and the threshold (Table~\ref{tab:resolution_scan}). At fixed threshold, increasing $k_B$ from 0.008 to 0.016 monotonically increases the resolution because the stronger quenching amplifies the relative effect of energy deposition fluctuations. At fixed $k_B$, lowering the threshold improves the relative resolution because the mean response increases more than its absolute event-by-event fluctuation.

\begin{table}[htbp]
\centering
\caption{Energy resolution $\sigma_{\rm dep}$ (\%) for 5.138~MeV alpha particles at different thresholds and Birks constants.}
\label{tab:resolution_scan}
\begin{tabular}{c|c|c|c}
\hline
sE$_{\rm min}$ (keV) & $k_B=0.008$ & $k_B=0.012$ & $k_B=0.016$ \\
\hline
0.25 & 1.27\% & 1.35\% & 1.39\% \\
0.5  & 1.42\% & 1.54\% & 1.61\% \\
1.0  & 1.65\% & 1.84\% & 1.96\% \\
\hline
\end{tabular}
\end{table}

The resolution improves with decreasing threshold at all $k_B$ values. For example, at $k_B=0.008$, lowering the threshold from 1.0 to 0.25~keV improves the resolution from 1.65\% to 1.27\%. Conversely, at fixed threshold, increasing $k_B$ from 0.008 to 0.016 worsens the resolution by 0.1--0.3 percentage points, with the largest effect at the highest threshold where the delta electron multiplicity is lowest.

\subsection{Variance Decomposition}
\label{sec:variance}

The total variance of the quenched light yield can be decomposed into primary and secondary contributions:

\begin{equation}
\mathrm{Var}(Q_{\text{total}}) = \mathrm{Var}(Q_{\text{primary}}) + \mathrm{Var}(Q_{\text{secondary}}) + 2\,\mathrm{Cov}(Q_{\text{primary}}, Q_{\text{secondary}}).
\label{eq:variance}
\end{equation}

\begin{figure}[htbp]
\centering
\includegraphics[width=0.7\textwidth]{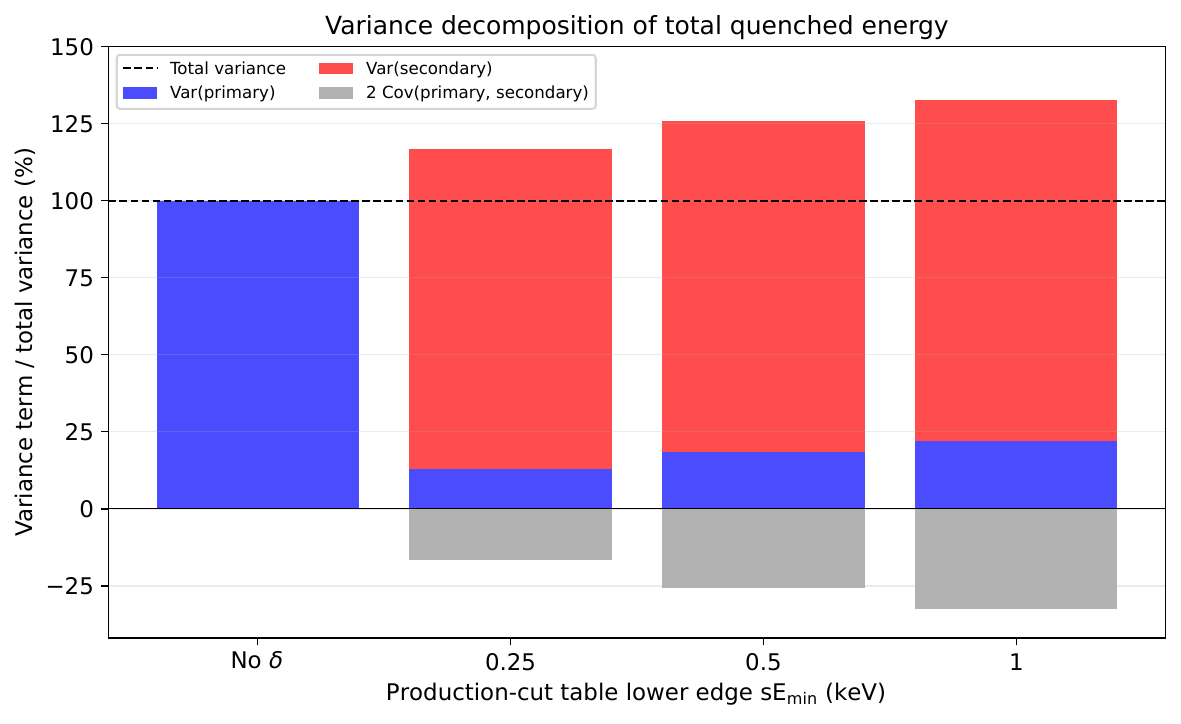}
\caption{Variance decomposition at 5.138~MeV and $k_B=0.008$ as function of delta electron threshold. The primary variance, secondary variance, and twice their covariance are normalized to the total variance and sum to 100\%.}
\label{fig:variance}
\end{figure}

The secondary (delta electron) contribution dominates at all thresholds (Fig.~\ref{fig:variance}), with its variance approximately equal to or larger than the total variance. The anti-correlation between primary and secondary qedep arises from energy conservation: when the alpha deposits more energy in the primary track, less is available for delta electrons, and vice versa. This anti-correlation reduces the total variance through the negative covariance term in Eq.~\ref{eq:variance}.

The dominance of the delta electron contribution can be understood from the event-by-event variation in the energy carried by delta electrons. The quenched response correlates strongly with the summed birth energy of direct delta electrons ($r=0.89$--0.94), rather than being determined by their multiplicity alone. Across independent runs over the tested energies, step limits, and thresholds, the normalized variance terms are 1.01--1.12 for the secondary contribution and 0.10--0.23 for the primary contribution, with the negative covariance reducing the total variance.

\subsection{Threshold and Physics Model Dependence}
\label{sec:threshold}

The delta electron production threshold is a critical parameter in the simulation. We scan the threshold at three values (sE$_{\rm min}=0.25$, 0.5, 1.0~keV) for 5.138~MeV alpha particles, comparing the Standard, Penelope, and Livermore electromagnetic models.

\begin{figure}[htbp]
\centering
\includegraphics[width=0.85\textwidth]{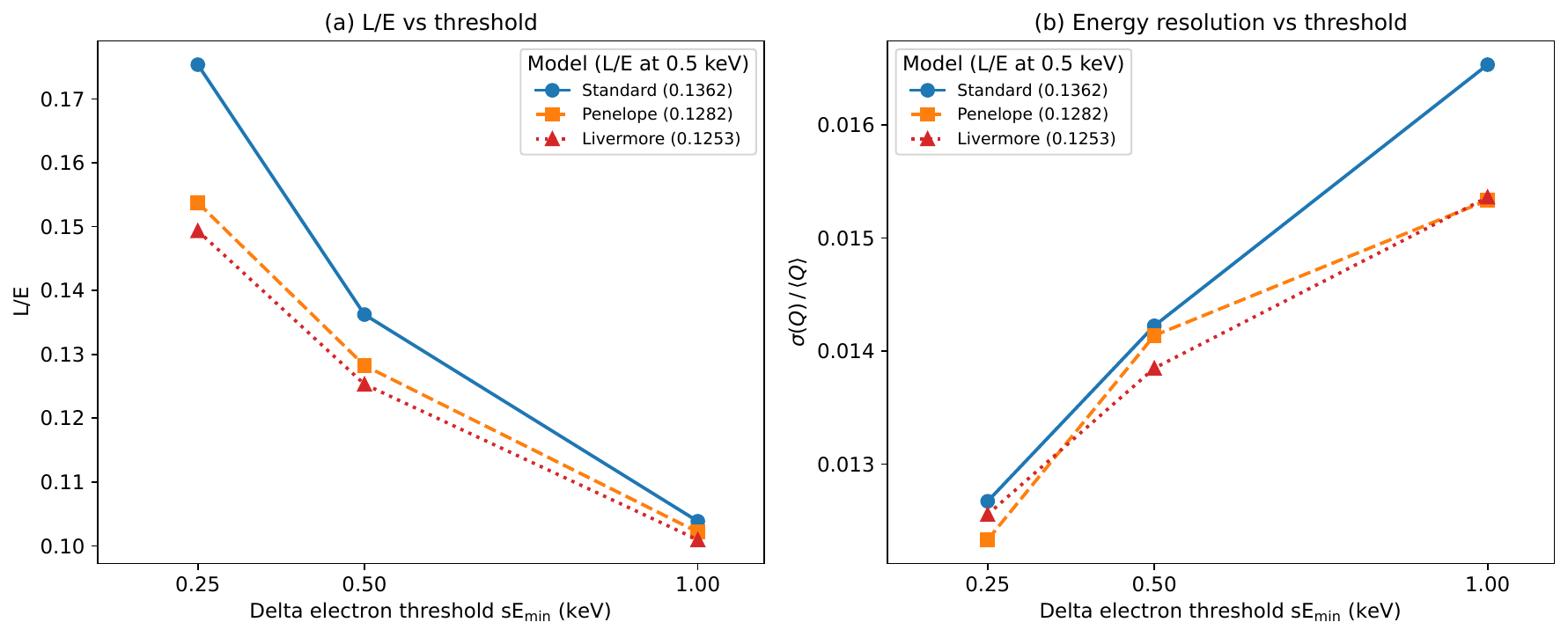}
\caption{Dependence of (a) L/E and (b) energy resolution $\sigma_{\rm dep}$ on the delta electron threshold for three electromagnetic models at 5.138~MeV.}
\label{fig:threshold}
\end{figure}

\begin{table}[htbp]
\centering
\caption{Threshold and model dependence for 5.138~MeV alpha particles. $N_\delta$ is the mean unique direct-electron count.}
\label{tab:threshold}
\begin{tabular}{c|c|c|c|c|c}
\hline
Model & sE$_{\rm min}$ (keV) & $N_\delta$ & $L/E$ & $\sigma_{\rm dep}$ & $p_{\rm frac}$ \\
\hline
Standard & 0.25 & 2346 & 0.1754 & 0.0127 & 0.402 \\
Standard & 0.5  & 820  & 0.1362 & 0.0142 & 0.552 \\
Standard & 1.0  & 207  & 0.1039 & 0.0165 & 0.744 \\
Penelope & 0.25 & 2349 & 0.1537 & 0.0123 & 0.458 \\
Penelope & 0.5  & 825  & 0.1282 & 0.0141 & 0.585 \\
Penelope & 1.0  & 210  & 0.1022 & 0.0153 & 0.754 \\
Livermore & 0.25 & 2348 & 0.1493 & 0.0126 & 0.472 \\
Livermore & 0.5  & 824  & 0.1253 & 0.0138 & 0.599 \\
Livermore & 1.0  & 210  & 0.1010 & 0.0154 & 0.764 \\
\hline
\end{tabular}
\end{table}

Lowering the threshold from 1.0 to 0.25~keV increases the delta electron multiplicity by a factor of $\sim$11, increases the L/E by a factor of $\sim$1.5--1.7, and improves the resolution by 0.27--0.39 percentage points. The primary fraction $p_{\rm frac}$ decreases from $\sim$0.75 to $\sim$0.40--0.47 as the threshold is lowered, confirming that the delta electron contribution dominates at low thresholds.

Penelope and Livermore agree with each other to within 2--3\% in L/E. The Standard model gives higher L/E than Penelope and Livermore, with the difference largest at the lowest threshold (about 14--17\% at 0.25~keV) and small at 1.0~keV (about 2--3\%). The threshold dependence is therefore larger than the difference among the tested configurations for the resolution, while the difference in absolute light yield remains non-negligible.

\section{Discussion}

\subsection{Origin of the Resolution Difference}

The alpha particle's intrinsic energy resolution ($\sigma_{\text{dep}} \approx 1.4\%$ at 5.138~MeV) is dominated by fluctuations in the delta electron population, as shown by the variance decomposition in Fig.~\ref{fig:variance}. The primary-track contribution has $\operatorname{Std}(Q_p)/\langle Q\rangle\approx0.61\%$, while the no-delta baseline at this energy is approximately 0.95\%. The negative primary--secondary covariance reduces the total variance.

This differs from gamma rays, where the energy is divided among Compton and photoelectric electrons with different quenching factors. Fluctuations in this energy partition can lead to a larger $\sigma_{\text{dep}}$ for gamma rays compared to alpha particles with the same total light yield.

\subsection{Step Size Effects and Implications for Simulations}

In the tested no-delta configuration, a user step limit changes the simulation from a deterministic single-step calculation to a multi-step calculation with a nonzero response width. Reducing the step limit from 0.02 to 0.01~$\mu$m changes the mean response by approximately 2--8\% over the tested energies, while the resolution changes by less than 3\%, indicating that its dependence on step size is already small over this interval.

The strong dependence of the alpha resolution on delta electron tracking implies that calibration sources emitting alpha particles require careful modeling of the energy deposition process. The production cut, step limiter, and effective delta electron threshold must be chosen consistently to avoid systematic biases in the simulated energy resolution.

\subsection{Challenges in Simulating keV Delta Electron Quenching}

The results in Sec.~\ref{sec:results} were obtained using the standard Geant4 physics with the Birks formula applied per step, at a nominal delta electron threshold of sE$_{\rm min}=0.5$~keV (Sec.~\ref{sec:threshold}). This approach assumes that the energy deposition of each delta electron can be approximated as a continuous straight track. However, for keV-scale electrons, this assumption may break down. We examine this issue using Geant4-DNA track structure simulations.

The simulation of keV-scale delta electron quenching faces several fundamental challenges related to the breakdown of the continuous track model at low electron energies.

\subsubsection{Track Structure at Low Energies: Geant4-DNA Results}

Condensed-history electromagnetic transport can represent keV electron energy loss with relatively coarse effective steps compared with the microscopic track structure. However, at these energies, the electron track is highly curved due to multiple scattering. We used Geant4-DNA with the $G4EmDNAPhysics\_option2$ physics list to study the track structure of low-energy electrons in liquid water. This water simulation is used as a qualitative proxy for track topology, not as a direct LAB material prediction.

\begin{figure}[htbp]
\centering
\includegraphics[width=0.85\textwidth]{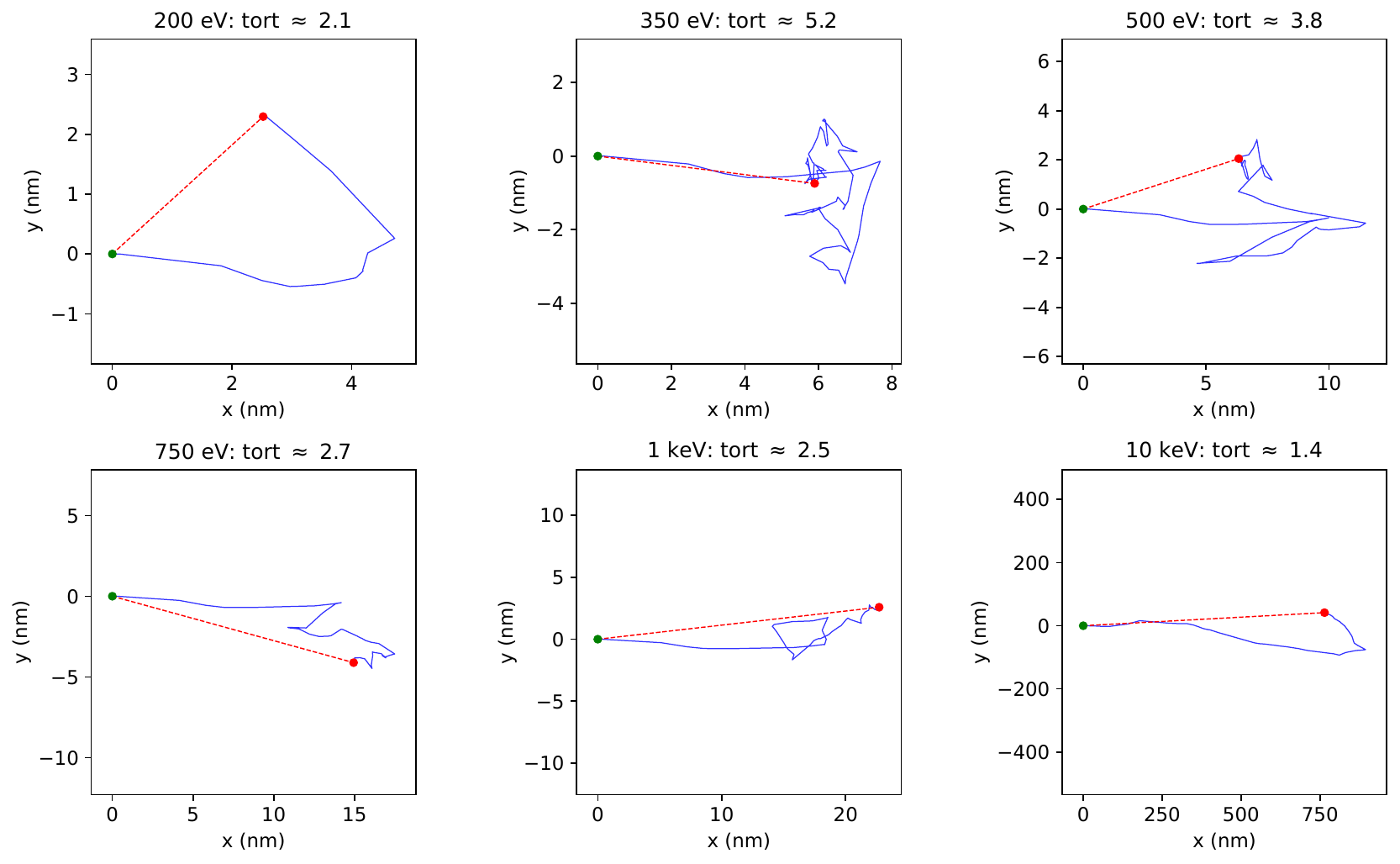}
\caption{Geant4-DNA simulation of electron tracks in water, shown in two-dimensional projection. The straight-line distance (red dashed line) is much shorter than the actual path length at low energies, with the sample mean tortuosity decreasing from $\sim$6 at 200~eV to $\sim$1.6 at 10~keV.}
\label{fig:dnatracks}
\end{figure}

The tortuosity (complete three-dimensional path length, divided by the initial-to-final distance) quantifies the curvature of the electron track:

\begin{table}[htbp]
\centering
\caption{Electron track tortuosity from Geant4-DNA.}
\label{tab:tortuosity}
\begin{tabular}{c|c|c|c|c}
\hline
Energy & Mean tortuosity & Median tortuosity & Mean path length & Mean straight distance \\
\hline
200~eV & $6.33 \pm 8.40$ & 4.16 & 30.1~nm & 5.9~nm \\
350~eV & $4.65 \pm 3.58$ & 3.56 & 39.2~nm & 10.0~nm \\
500~eV & $3.53 \pm 2.54$ & 2.70 & 47.5~nm & 15.4~nm \\
750~eV & $2.99 \pm 3.00$ & 2.30 & 68.5~nm & 27.4~nm \\
1~keV & $2.51 \pm 1.31$ & 2.15 & 89.4~nm & 40.1~nm \\
10~keV & $1.57 \pm 0.57$ & 1.41 & 2725.4~nm & 1856.3~nm \\
\hline
\end{tabular}
\end{table}

\begin{figure}[htbp]
\centering
\includegraphics[width=0.65\textwidth]{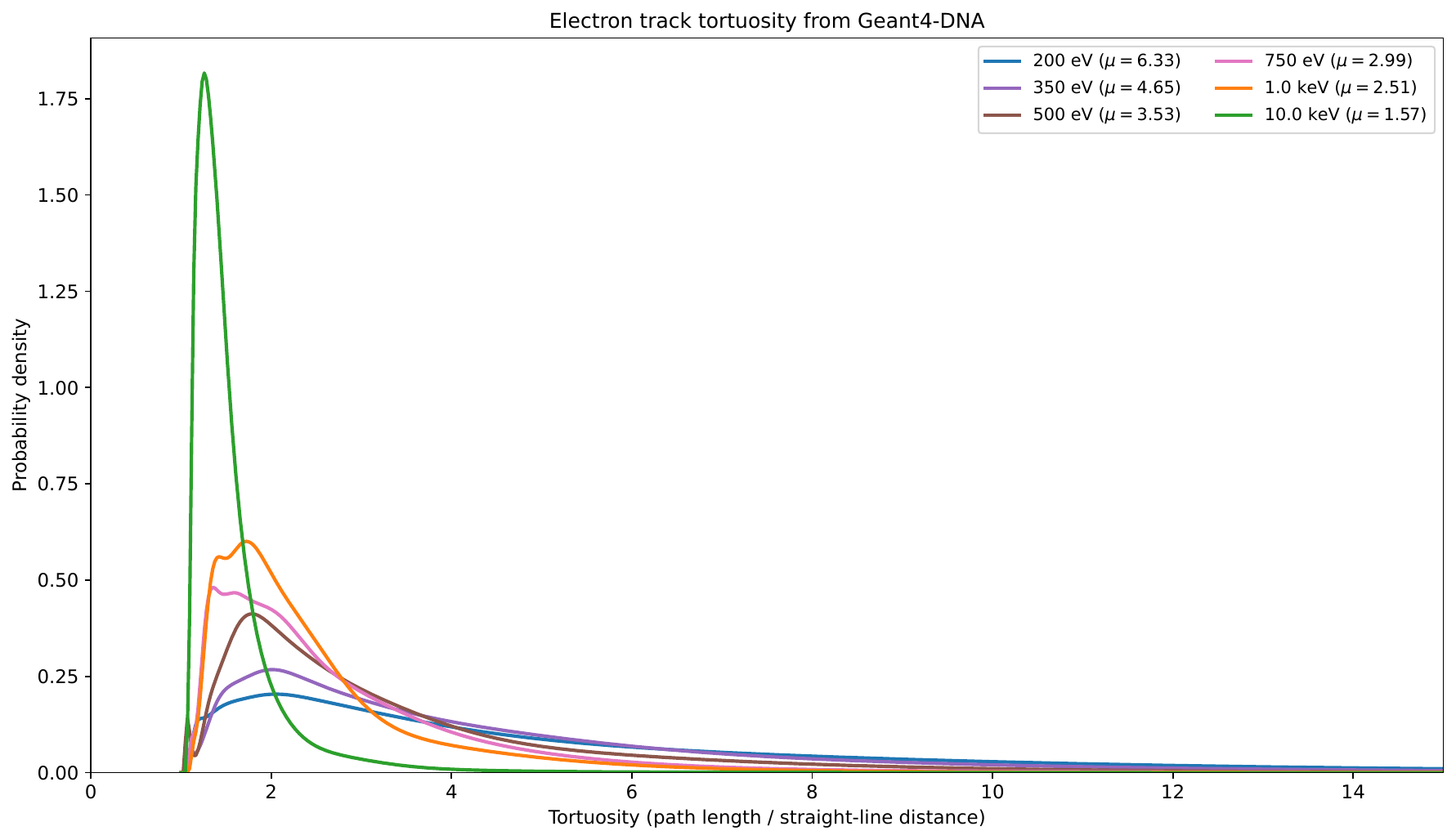}
\caption{Distribution of electron track tortuosity from Geant4-DNA at six energies from 200~eV to 10~keV. The distribution is broad and asymmetric, with a long tail at high tortuosity for low energies.}
\label{fig:tortuosity}
\end{figure}

The tortuosity increases dramatically at low energies: at 200~eV, the median tortuosity is 4.16, meaning the actual path is nearly 4 times longer than the straight-line distance. The mean tortuosity is even higher (6.33) due to rare events with extreme curvature (tortuosity $>$ 20). The energy is deposited in a small ``blob'' ($\sim$5~nm radius for 200~eV), not a continuous track. The tortuosity decreases monotonically with energy: 4.65 at 350~eV, 3.53 at 500~eV, 2.99 at 750~eV, 2.51 at 1~keV, and 1.57 at 10~keV. The Birks formula, which assumes a continuous cylindrical column, may not be valid for such energy deposition topologies.

The distribution of tortuosity (Fig.~\ref{fig:tortuosity}) is broad and asymmetric, following a gamma distribution. This is important because the quenching factor depends on the local dE/dx, which is affected by the actual path length: a more tortuous path spreads the energy over a larger volume, reducing the local ionization density and therefore the quenching.

\subsubsection{Blob Quenching vs.\ Column Quenching}

For sub-keV electrons, the Mozumder-Magee classification~\cite{mozumder} categorizes energy deposition as ``spurs'' ($<$100~eV, $\sim$2--4~nm radius) and ``blobs'' (100--500~eV, $\sim$5--8~nm radius). The Onsager radius in organic scintillators ($\sim$28~nm) -- the distance at which the Coulomb binding energy equals the thermal energy -- is much larger than these spur/blob sizes. This means that electron-ion pairs within a blob are strongly bound and geminate recombination is nearly complete, but the competition between radiative and non-radiative recombination within the blob is not well understood.

The Christensen-Andersen ExcitonQuenching model~\cite{christensen}, developed for organic plastic scintillators (polystyrene and polyvinyltoluene), addresses this by solving the 3D reaction-diffusion equation (Blanc equation) with the radial energy density distribution from track structure theory. However, this model is currently limited to heavy particles (protons, alphas) with cylindrical track geometry and does not support electrons with spherical spur/blob topology.

\subsubsection{Core vs.\ Penumbra Contribution to Alpha Light Yield}

The ExcitonQuenching model was developed for organic plastic scintillators (BCF-12, BC-400, based on polystyrene and polyvinyltoluene) and validated against proton quenching data~\cite{christensen}. It separates the ion track into a high-density core (radius $\sim$10--15~nm) and a penumbra dominated by delta electrons, with exciton density falling as $1/r^2$. The model solves the Blanc equation for the spatio-temporal evolution of excitons, including diffusion, fluorescence, and bimolecular quenching.

Applying this model to LAB (light yield 10000~ph/MeV, decay time 3.5~ns, density 0.859~g/cm$^3$) with the Scholz-Kraft track structure and the quenching parameters fitted to plastic scintillator data ($\alpha = 9\times10^{-8}$~cm$^3$/s, $D = 5\times10^{-4}$~cm$^2$/s) gives a striking result for an 8~MeV alpha particle: approximately 89\% of the scintillation light originates from the penumbra (delta electrons), and only $\sim$11\% from the core. This is because the core exciton density is so high ($\sim$10$^{17}$~cm$^{-3}$) that bimolecular quenching destroys almost all excitons before they can fluoresce. The penumbra, with its much lower density, has a survival fraction of $\sim$27\%.

This prediction is far more extreme than our Geant4+Birks result, which gives approximately equal contributions from the core and delta electrons. This discrepancy should be interpreted as model-dependent rather than as a validated physical prediction for LAB. One possible reason is that the ExcitonQuenching model's $\alpha$ parameter was fitted to plastic scintillator data, where the solid polymer matrix restricts molecular motion. In liquid scintillators, different diffusion and bimolecular-quenching parameters could change the core contribution substantially.

The Birks formula itself can be seen as a simplified solution to the Blanc equation~\cite{christensen}, where the diffusion and bimolecular quenching are subsumed into the single parameter $k_B$. The SNO+ comparison constrains only the broad alpha-quenching shape within sizable uncertainties and cannot determine the core-penumbra balance uniquely. Determining the $\alpha$ and $D$ parameters independently for LAB would require dedicated quenching measurements with monoenergetic electrons and ions, which are currently unavailable.

\subsubsection{Absence of Experimental Data}

There are no direct experimental measurements of the scintillation efficiency of monoenergetic sub-keV electrons in organic liquid scintillators. The stopping power of electrons in the 0.1--10~keV range has an uncertainty of 10--30\% depending on the model (Penelope vs.\ Bethe-Bloch), and this uncertainty propagates directly to the quenching calculation.

Furthermore, there are no experimental measurements of the alpha particle energy-dependent quenching curve in liquid scintillators with a precision better than 1\%, nor any measurements of the intrinsic alpha energy resolution with a precision better than 2\%. The SNO+ data~\cite{vonkrosigk} used in this work has relative uncertainties of 3--5\% on the quenching factor. This lack of high-precision data prevents a definitive validation of the core vs.\ penumbra light partition predicted by different models.

\subsubsection{Possible Solutions}

Several approaches could address these challenges:

\begin{enumerate}
\item \textbf{Energy-dependent quenching table}: Construct a table of $Q_{\text{avg}}(E)$ for electrons by running Geant4 simulations at each energy and applying the step-by-step Birks formula. This table can then be used in a separate simulation to replace the per-step Birks calculation for delta electrons, ensuring consistency.

\item \textbf{Geant4-DNA per-step quenching}: Use Geant4-DNA track structure (which correctly models the curved path of keV electrons) and apply Birks quenching to each step. This is computationally expensive but physically more accurate.

\item \textbf{ExcitonQuenching extension}: Extend the Christensen-Andersen model to spherical geometry for electron blobs, using the radial energy density distribution from electron track structure theory. This would provide a first-principles calculation of sub-keV electron quenching.

\item \textbf{Experimental measurements}: Direct measurements of the scintillation light yield from monoenergetic sub-keV electrons (e.g., using a proportional counter or photoemission source) would provide crucial validation data for the models.
\end{enumerate}

\section{Conclusions}

We have presented a detailed Geant4 simulation study of alpha particle energy deposition and light yield in LAB-based liquid scintillators, motivated by the increasing use of Po isotope alpha particles for energy calibration in large LS detectors.

Without a step limit, Geant4 simulates the alpha in a single CSDA step, yielding a deterministic result ($\sigma=0$). A step limit is required to introduce stochastic energy loss fluctuations, producing the physical range straggling ($\sigma\sim1\%$). The effective delta electron threshold is controlled by the interplay of the Bragg model's $E_{\rm min}$ parameter and the production cut; the default $E_{\rm min}=0.25$~keV/nucleon clamps the threshold at 0.99~keV regardless of the cut.

Delta electrons carry 18--20\% of the alpha's energy but contribute 48--50\% of the quenched light yield, because their lower dE/dx results in weaker Birks quenching. The simulation is compatible with the digitized SNO+ alpha quenching curve shape within current uncertainties, with $\chi^2$/ndf $\simeq0.14$--0.17 for $k_B=0.007$--0.010~g$\cdot$cm$^{-2}\cdot$MeV$^{-1}$. The no-delta case gives a worse shape comparison ($\chi^2$/ndf $\simeq0.24$ at $k_B=0.008$), supporting the essential role of delta electrons.

The intrinsic energy resolution ($\sigma_{\rm dep}=1.42\%$ at 5.138~MeV) is dominated by delta electron fluctuations, which account for $\sim$95\% of the total variance. The resolution depends on both $k_B$ and the threshold, with the threshold setting being the dominant systematic uncertainty in the Geant4 11.4.2 study. Penelope and Livermore agree with each other within 2--3\% in L/E, while the Standard model gives moderately higher light yield, especially at low thresholds.

Geant4-DNA simulations show that keV-scale electron tracks are highly curved (median tortuosity 4.16 at 200~eV), challenging the continuous column approximation for Birks quenching. The ExcitonQuenching model predicts that $\sim$89\% of the alpha scintillation light originates from the delta electron penumbra, though this extreme ratio requires experimental validation. High-precision measurements of alpha quenching and resolution in liquid scintillators are needed to further constrain the simulation parameters.

\section*{Declarations}

\textbf{Funding} The work is supported by the National Key R\&D Program of China under Grant No.~2023YFA1606102.

\textbf{Competing interests} The authors have no competing interests to declare that are relevant to the content of this article.

\textbf{Data availability} The simulation output files and analysis scripts are available from the corresponding author upon reasonable request.

\end{document}